\documentclass[twoside,twocolumn,9pt]{article}
\usepackage{extsizes}
\usepackage[super,sort&compress,comma]{natbib} 
\usepackage[version=3]{mhchem}
\usepackage[left=1.5cm, right=1.5cm, top=1.785cm, bottom=2.0cm]{geometry}
\usepackage{balance}
\usepackage{mathptmx}
\usepackage{sectsty}
\usepackage{graphicx} 
\usepackage{lastpage}
\usepackage[format=plain,justification=justified,singlelinecheck=false,font={stretch=1.125,small,sf},labelfont=bf,labelsep=space]{caption}
\usepackage{float}
\usepackage{fancyhdr}
\usepackage{fnpos}
\usepackage[english]{babel}
\addto{\captionsenglish}{%
  
}
\usepackage{array}
\usepackage{droidsans}
\usepackage{charter}
\usepackage[T1]{fontenc}
\usepackage[usenames,dvipsnames]{xcolor}
\usepackage{setspace}
\usepackage[compact]{titlesec}
\usepackage{hyperref}
\usepackage{soul}

\soulregister{\cite}{7}
\soulregister{\comment}{7}

\newcommand{\comment}[1]{{\color{red}#1}}

\usepackage{epstopdf}

\definecolor{cream}{RGB}{222,217,201}

\begin{document}

\pagestyle{fancy}
\thispagestyle{plain}
\fancypagestyle{plain}{
\renewcommand{\headrulewidth}{0pt}
}

\makeFNbottom
\makeatletter
\renewcommand\LARGE{\@setfontsize\LARGE{15pt}{17}}
\renewcommand\Large{\@setfontsize\Large{12pt}{14}}
\renewcommand\large{\@setfontsize\large{10pt}{12}}
\renewcommand\footnotesize{\@setfontsize\footnotesize{7pt}{10}}
\makeatother

\renewcommand{\thefootnote}{\fnsymbol{footnote}}
\renewcommand\footnoterule{\vspace*{1pt}%
\color{cream}\hrule width 3.5in height 0.4pt \color{black}\vspace*{5pt}} 
\setcounter{secnumdepth}{5}

\makeatletter 
\renewcommand\@biblabel[1]{#1}            
\renewcommand\@makefntext[1]%
{\noindent\makebox[0pt][r]{\@thefnmark\,}#1}
\makeatother 
\renewcommand{\figurename}{\small{Fig.}~}
\sectionfont{\sffamily\Large}
\subsectionfont{\normalsize}
\subsubsectionfont{\bf}
\setstretch{1.125} 
\setlength{\skip\footins}{0.8cm}
\setlength{\footnotesep}{0.25cm}
\setlength{\jot}{10pt}
\titlespacing*{\section}{0pt}{4pt}{4pt}
\titlespacing*{\subsection}{0pt}{15pt}{1pt}

\fancyfoot{}
\fancyfoot[LO,RE]{\vspace{-7.1pt}\includegraphics[height=9pt]{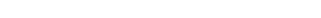}}
\fancyfoot[CO]{\vspace{-7.1pt}\hspace{13.2cm}\includegraphics{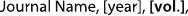}}
\fancyfoot[CE]{\vspace{-7.2pt}\hspace{-14.2cm}\includegraphics{head_foot/RF}}
\fancyfoot[RO]{\footnotesize{\sffamily{1--\pageref{LastPage} ~\textbar  \hspace{2pt}\thepage}}}
\fancyfoot[LE]{\footnotesize{\sffamily{\thepage~\textbar\hspace{3.45cm} 1--\pageref{LastPage}}}}
\fancyhead{}
\renewcommand{\headrulewidth}{0pt} 
\renewcommand{\footrulewidth}{0pt}
\setlength{\arrayrulewidth}{1pt}
\setlength{\columnsep}{6.5mm}
\setlength\bibsep{1pt}

\makeatletter 
\newlength{\figrulesep} 
\setlength{\figrulesep}{0.5\textfloatsep} 

\newcommand{\topfigrule}{\vspace*{-1pt}%
\noindent{\color{cream}\rule[-\figrulesep]{\columnwidth}{1.5pt}} }

\newcommand{\botfigrule}{\vspace*{-2pt}%
\noindent{\color{cream}\rule[\figrulesep]{\columnwidth}{1.5pt}} }

\newcommand{\dblfigrule}{\vspace*{-1pt}%
\noindent{\color{cream}\rule[-\figrulesep]{\textwidth}{1.5pt}} }

\makeatother

\twocolumn[
  \begin{@twocolumnfalse}
{\includegraphics[height=30pt]{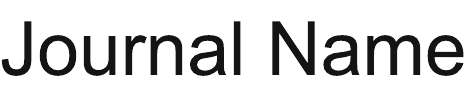}\hfill\raisebox{0pt}[0pt][0pt]{\includegraphics[height=55pt]{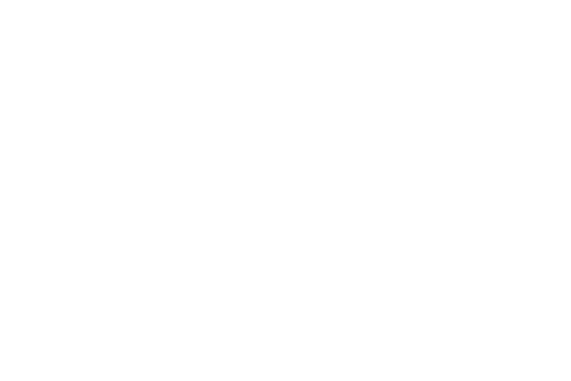}}\\[1ex]
\includegraphics[width=18.5cm]{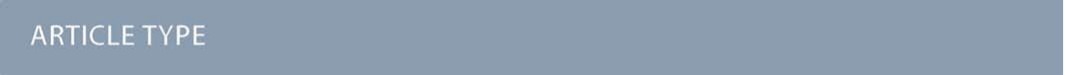}}\par
\vspace{1em}
\sffamily
\begin{tabular}{m{4.5cm} p{13.5cm} }

\includegraphics{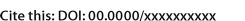} & \noindent\LARGE{\textbf{A common control and data layer for automated electron microscopy}} \\
\vspace{0.3cm} & \vspace{0.3cm} \\

 & \noindent\large{Cedric Lim\textit{$^{a}$}, Alexander J. Pattison\textit{$^{b}$}, Peter Ercius\textit{$^{c}$}, Mary C. Scott\textit{$^{c,d}$}, Jörg Tatchen\textit{$^{e}$}, Kersten Hahn\textit{$^{e}$}, Peter A. van Aken\textit{$^{e}$}, Karen Bustillo\textit{$^{c}$}, Colin Ophus\textit{$^{a}$} and Tara P. Mishra\textit{$^{e\ddag}$}} \\

\includegraphics{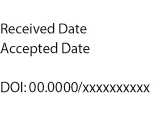} & \noindent\normalsize{

Self-driving laboratories can now synthesize materials faster than they can be characterized. Transmission electron microscopy is one of the most informative characterization techniques, but remains among the hardest to automate. Each vendor exposes its own scripting interface, protection of the instrument is rebuilt for every new script or agent, and the conditions of each measurement are usually reconstructed after the experiment. Here we present NIMBUS, an open platform that controls JEOL and Thermo Fisher Scientific microscopes through a single generalized API, using only a small server on the microscope computer. Before an experiment starts, NIMBUS checks that the instrument provides the capabilities it needs, and that instrument-specific guardrails protect the microscope during unattended operation. In the experiments undertaken using the NIMBUS platform, each acquisition is written directly to a community-standard NXem record. These records include an explicit list of what the instrument could not report. The same code runs a correlated imaging, X-ray spectroscopy and diffraction workflow on microscopes from two different vendors, and drives an unattended electron tomography acquisition. NIMBUS provides a common foundation for automated electron microscopy in which every measurement leaves an interpretable record and a unified basis on which agentic control can be built.
}
\end{tabular}

 \end{@twocolumnfalse} \vspace{0.6cm}

  ]

\renewcommand*\rmdefault{bch}\normalfont\upshape
\rmfamily
\section*{}
\vspace{-1cm}


\footnotetext{\textit{$^{a}$~Department of Materials Science and Engineering, Stanford University, Stanford, CA 94305, USA}}
\footnotetext{\textit{$^{b}$~Materials Characterization and Processing Facility, Whiting School of Engineering, Johns Hopkins University, Baltimore, MD 21218, USA}}
\footnotetext{\textit{$^{c}$~National Center for Electron Microscopy, Molecular Foundry, Lawrence Berkeley National Laboratory, Berkeley, CA 94720, USA}}
\footnotetext{\textit{$^{d}$~Department of Materials Science and Engineering, University of California, Berkeley, CA 94720, USA}}
\footnotetext{\textit{$^{e}$~Max Planck Institute for Solid State Research, Stuttgart 70569, Germany}}


\footnotetext{\ddag: Corresponding author: Tara P. Mishra, email: t.mishra@fkf.mpg.de}




\section{Introduction}

Advances in reinforcement learning~\cite{qiu2026keyframe, huang2025mobile} and agentic workflows~\cite{anthropic2026enzyme, angelopoulos2026prompts, jiang2026selfevolving, kusne2026managing} are reshaping research in biology, chemistry and materials science, with self-driving laboratories now planning, executing, and interpreting sequences of experiments with little human intervention. However, sample throughput has begun to outpace the rate at which those samples can be characterized~\cite{szymanski2023alab, burger2020mobile, macleod2020ada}. This is especially true in materials science, where advanced characterization at the nanoscale still depends on an expert operator, even when the rest of the laboratory is automated. The characterization step has therefore become the bottleneck that limits the pace of materials discovery~\cite{kalinin2021automated}.

Among these techniques, transmission electron microscopy (TEM) is one of the most informative and one of the most difficult to automate. A single instrument can acquire images, diffraction patterns and chemical spectra from the same region of a specimen, resolving the structure and composition that determine material properties at the nanoscale. However, the electron beam can alter the specimen during acquisition, and the instrument conditions can drift between sessions and vary between operators, so a measurement can rarely be repeated under identical conditions. Every TEM dataset is therefore costly to obtain, difficult to reproduce, and only as informative as the record of how it was acquired.

To address this, there have been efforts to automate TEM-based characterization. In the cryo-EM community, frameworks such as SerialEM have long been used to automate data collection~\cite{mastronarde2003serialem, mastronarde2005automated}. For materials applications, vendors provide programmable interfaces, such as PyJEM for JEOL microscopes and TEM Scripting, TEM Imaging and Analysis (TIA) and AutoScript for Thermo Fisher microscopes. Open-source packages such as Instamatic and ST4DeM build on these interfaces to provide instrument abstraction, remote control and automated acquisition~\cite{cichocka2018instamatic, smeets2021instamatic, uusimaki2025st4dem}, and stemOrchestrator, now continued as Asyncroscopy, combines the interfaces of the microscope, the aberration corrector and the spectrometer on a single Thermo Fisher column into one framework for automated workflows~\cite{pratiush2025orchestrator, houston2025asyncroscopy}. Machine-learning-driven acquisition has further been used to correct aberrations, select regions of interest and choose the next measurement based on the previous one~\cite{pattison2025beacon, roccapriore2022physics, pratiush2025hae, pratiush2025seek}. 

More recently, agentic frameworks have been applied to electron microscopy. Large language models (LLMs) can already analyze experimental data and make data-driven decisions, but giving them full control of a microscope risks expensive damage to the instrument or a hallucinated decision path that can lead to an incorrect experiment~\cite{li2024enhancing, mandal2025evaluating, guinan2025mind}. Exposing the full scripting interfaces of several vendors to an agent would also spend much of the model's limited context on instrument details rather than on the experiment itself. To address this, a platform by Borch \textit{et al.}, ADAM (Autonomous Decision-Making Agent for Material Experimentation)~\cite{spurgeon2026balancing} limits hallucinations through a multi-agent framework with a model context protocol (MCP) server that offers a restricted set of safe microscope controls and operating documentation, with a human checking key decisions. TEM Agent~\cite{wall2026tem} similarly connects a commercial LLM through MCP servers to a Thermo Fisher microscope, to the aberration corrector, to the Crucible data management platform at the Molecular Foundry and to high performance computing resources. Recently, a related approach by Slautin \textit{et al.} in scanning probe microscopy restricts the decisions given to the agent rather than the controls it can reach. In this approach, an agent decides which measurement is needed next, while deterministic algorithms compute the exact measurement position and execute the command on the instrument~\cite{slautin2026hierarchical}.

Platforms like these show the frontier of automated microscope control, but they also leave open questions and gaps. For example, how costly agentic workflows are in practice, and how far conventional algorithmic acquisition can still be improved, remain unclear. As open-source language models now trail the best commercial models by only a few months~\cite{epoch2026openclosed, bengio2026international}, attention is shifting to the software around the model, often called the harness, which determines which tools an agent can use and what information it receives. For a microscope, this harness has to include structured control of the instrument. Today, however, the protection of the microscope is defined separately in each framework, and each implementation is built around the microscope interface of one manufacturer, so a workflow developed in one laboratory can rarely be reused on an instrument from another. The record of each measurement is similarly fragmented. In an unsupervised autonomous run, the record is the only account of what was done by the instrument. Yet where current frameworks manage data, they do so through the platform of their own facility rather than in a representation shared across instruments and laboratories. A common control layer with the protection of the instrument built in, together with a universal record written at acquisition, would let human experts and automated workflows work on the same data, with the expert able to follow, check and correct what was done.

\begin{figure*}[h]
\centering
  \includegraphics[width=\textwidth]{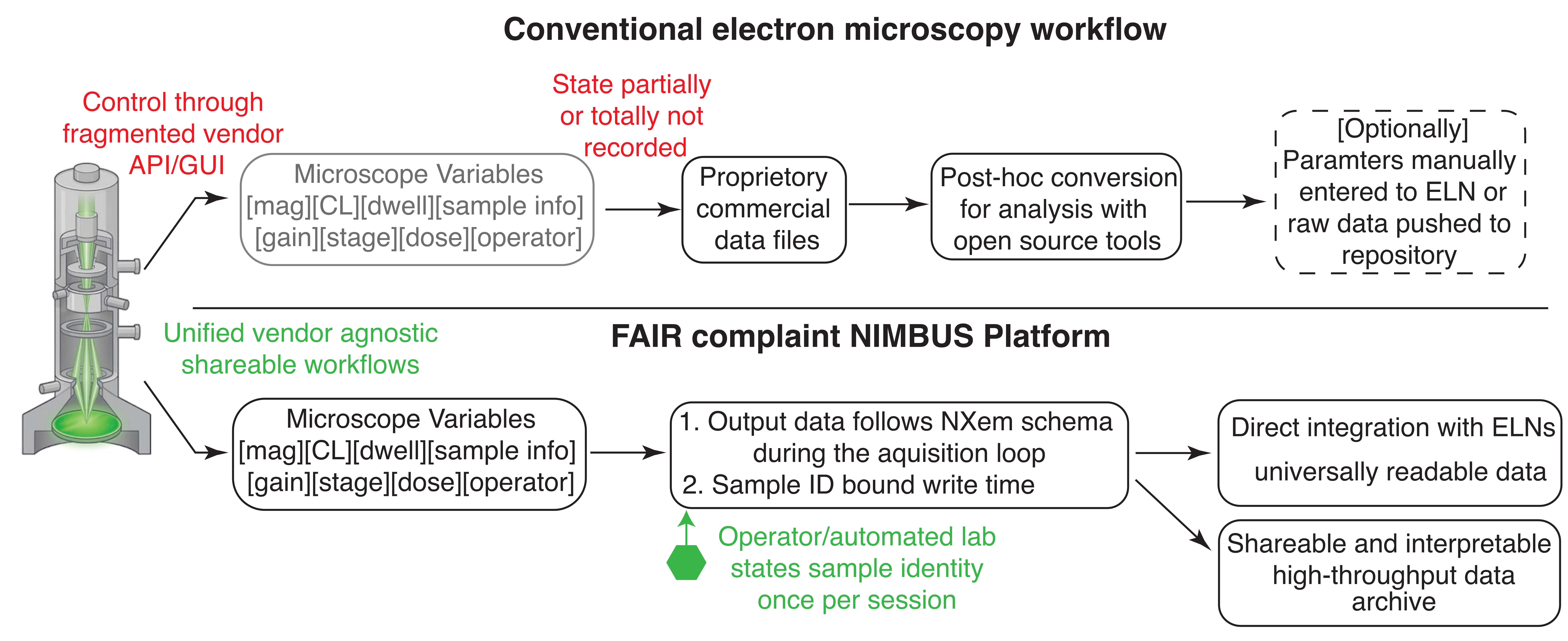}
  \caption{Conventional post-acquisition conversion (top) compared with acquisition-time recording through NIMBUS (bottom). Both paths begin at the same measurement, with the microscope holding the state that describes each signal such as magnification, camera length, dwell time, detector gain, stage position, high tension. The conventional path writes a vendor file and maps it to a community schema afterwards; the faded variables mark state that does not reach the standard record, which is a limit on what a converter can recover rather than a claim that vendor files record nothing. Optionally, the gap is filled by a manually completed electronic laboratory notebook file, once per conversion. The software package introduced in this work, NIMBUS, writes the NXem record inside the acquisition loop while that state is still available, leaving the specimen declaration as the only human input, once per session. A notebook (dashed) is required in the first path and optional in the second, where the archive is interoperable and reusable before any catalog exists and a notebook adds findability and access. Hexagons mark inputs about the investigated sample by a researcher or an automated lab.}
  \label{fgr:workflow}
\end{figure*}

The microscopy community has recognized the need for such a systematic data management and record. The MaRDA working groups recommend comprehensive metadata capture, together with standardized vocabularies and community identifiers~\cite{taillon2025marda}. The infrastructure to represent data in such a schema now exists in the ratified NeXus application definition for electron microscopy, NXem~\cite{konnecke2015nexus, nxem}, and in the persistent identifiers of the Helmholtz Metadata Collaboration EM Glossary~\cite{emglossary2024}. These resources define how metadata should be represented, but not when or how they should be captured. Converting vendor files afterwards, as pynxtools-em does, requires information to be added by hand from an electronic laboratory notebook, because parts of the instrument state are missing from the vendor output~\cite{pynxtools-em, shabih2026pynxtools} (Fig.~\ref{fgr:workflow}). At large-scale facilities, NeXus records are instead written during acquisition, for example by GDA at Diamond, Sardana at ALBA and Bluesky-based systems at BESSY II~\cite{gibbons2011gda, reszela2019generic, smith2022status, allan2019bluesky}. Transferring this approach to a laboratory microscope is not straightforward, since the microscope computer runs vendor-supported software, may have no external network access and often does not permit additional packages to be installed. 

With both automated control and data management in mind, we present NIMBUS (Next-gen Infrastructure for Microscopy and Built-in Unified Solutions for Analysis), a platform that controls JEOL and Thermo Fisher Scientific (TFS) microscopes through a single API and records every acquisition in a community standard. Since vendor microscope computers often cannot have new software installed, NIMBUS places only a small server on them and runs experiments from a separate workstation, a server model introduced by Pattison \textit{et al.} for TFS instruments~\cite{pattison2024advanced}. An earlier release, NIMBUS v1.0.0~\cite{mishra2026nimbusv1}, used this architecture without a vendor-agnostic control language or data layer. Even this initial platform has already proved useful in material science research. High-throughput automated microscopy using NIMBUS v1.0.0 traced an orthorhombic LiMnO$_2$ impurity phase in Mn-rich disordered rocksalt cathodes to interparticle compositional heterogeneity~\cite{lee2026near}.

In this manuscript, we introduce a substantially extended version of NIMBUS. Before an experiment starts, NIMBUS checks that the instrument provides the required capabilities, and instrument-specific guardrails protect the microscope during unattended operation, whether commands come from a script, an algorithm or an agent. Each acquisition is written into an NXem record with the instrument state and an explicit list of the quantities the instrument could not report, with no manual notebook entry (Fig.~\ref{fgr:workflow}). We demonstrate NIMBUS on a JEOL JEM-F200 at the Max Planck Institute for Solid State Research and a TFS ThemIS at the Molecular Foundry, Lawrence Berkeley National Laboratory, with a correlated imaging, X-ray energy-dispersive spectroscopy (EDS) and diffraction workflow run with the same code on both instruments, and an unattended electron tomography acquisition. We envision NIMBUS as a common foundation for algorithmic and agentic workflows in electron microscopy.

\section{Architecture and design principles}

\subsection{Architecture}

\begin{figure*}[h]
    \centering
    \includegraphics[width=0.95\textwidth]{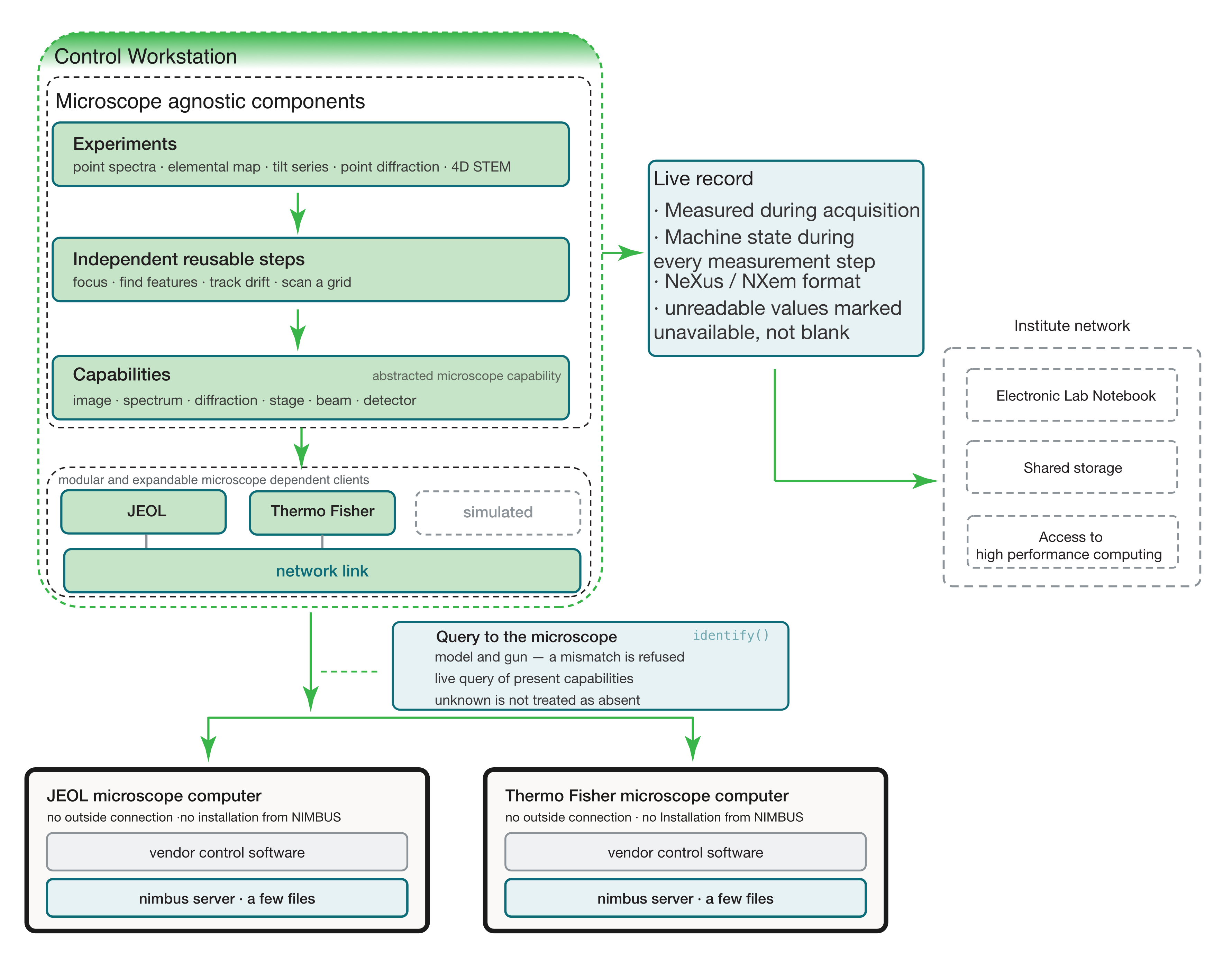}
    \caption{NIMBUS deployment. Experiments, reusable workflow steps and capabilities run on a control workstation and reach each microscope through a vendor-specific client and a small NIMBUS server on the microscope computer. The client checks the microscope model and its available capabilities before an experiment, and every acquisition is written to a live NXem record that can be passed to institute storage, an electronic lab notebook or computing resources.}
    \label{fig:fig2_deployment}
\end{figure*}

Each vendor exposes microscope capabilities through different scripting interfaces, and each differs in syntax and supported Python versions. JEOL microscopes are controlled through PyJEM and FEMTUS, while TFS microscopes use the TEM Scripting COM interface for column control and TIA or Velox for acquisition. Newer TFS microscopes also offer the AutoScript interface, which is not available on older platforms. One of the limitations is that these interfaces must run on the microscope computer itself, which is often offline. Therefore the scripts have to be written for, and run on, that specific machine. This prevents a script written for one instrument from being reused on another, even if they share the same manufacturer. In this work, NIMBUS therefore uses a server-client design. A lightweight Python server is copied onto the microscope computer rather than installed, wraps the vendor interface and exposes the microscope controls over the network. The client, running on a separate workstation, provides a generalized API for controlling the microscope and handles all experimental logic, metadata capture and NXem writing, communicating with the server over ZeroMQ on a private network link (Fig.~\ref{fig:fig2_deployment}). For each vendor, the client contains a thin driver that translates the generalized API into requests to the server. Experiments are written only against the generalized API and never call vendor-specific code.

The generalized API of NIMBUS is built from a set of capabilities, each describing an operation a microscope can perform. For example, acquiring a HAADF-STEM image requires different calls in PyJEM and in TEM Scripting and TIA, and NIMBUS abstracts these to expose a single \texttt{acquire\_haadf} function. Experimental workflows are written using this API and are therefore not tied to a single vendor and can be shared between instruments. However, a shared API does not mean each microscope can perform every operation, since microscopes are often fitted with different detectors with varying capabilities. To account for this, NIMBUS queries the instrument to determine the available capabilities of the microscope, rather than relying on a stored configuration. This handles cases where microscope parameters are readable in one microscope and not the other, and also prevents experiments from running without the required capabilities, rather than failing during runtime. 

NIMBUS is organized in three layers built on top of these capabilities. Capabilities are single operations, such as moving the stage, setting the defocus or acquiring a HAADF-STEM image. Workflows combine capabilities into routines that are useful on their own and recur across many experiments, such as finding the eucentric height, focusing, or aligning the image after a stage move. Experiments then combine workflows and capabilities into a complete measurement, for example a tilt series or a survey of particles, and define what is acquired and in what order. As each layer uses only the one below, a workflow can be improved or replaced without changing the experiments that use it, and a new experiment can reuse existing workflows rather than reimplementing them (Section~\ref{sec:tomography}). All experiments also record their data in the same way. Before each acquisition the experiment reads the state of the instrument in a single request and passes it, together with the data, to one recording layer shared by all experiments. The contents of this record and the handling of values that the instrument cannot report are described in Section~\ref{sec:record_instrument}.

Since the experimental workflows never call vendor-specific code, NIMBUS can be extended to new instruments without changing the workflows. We have written one base class for each vendor, which implements the generalized API, and a small class for each microscope model, which holds only what is particular to that instrument, such as detector names, image orientation and calibrations. On JEOL microscopes, the base class uses PyJEM and FEMTUS. Because PyJEM cannot report which microscope it is connected to, the model is declared when the server is started. The client then compares this declaration with the gun type read from the instrument and refuses the connection if they do not match, since a mislabeled session would write subtly wrong metadata into every record it produces. When the client connects, the JEOL server also reports which detectors and capabilities are present, so that an experiment needing missing hardware is refused before it starts. Models currently supported are the JEM-F200, tested on the instrument, and the ARM200F, whose differences, including a six-axis stage and an accessible gun current, are taken from the PyJEM documentation.

On TFS microscopes, the base class uses TEM Scripting for column control and TIA for acquisition, through a server on the microscope computer as described above. Newer TFS microscopes also ship with AutoScript, the vendor's own Python interface, which can be reached over the network. NIMBUS treats AutoScript as a second way of connecting to the same TFS class rather than as a separate microscope class, so every workflow and experiment runs over either connection without change. When using AutoScript, no NIMBUS server is needed on the microscope computer. The two connections can also be combined. On the Talos microscope at Stanford, X-EDS acquired through TIA returns spectra with no counts, while the same detector records correctly through AutoScript. The user therefore starts the server with AutoScript selected for X-EDS, while the HAADF image and the beam positions are still handled through TIA. The choice is made explicitly when the server is started and is never taken automatically. Where AutoScript cannot answer a request in the same form as TEM Scripting, NIMBUS refuses it rather than returning a substitute.

\subsection{Guardrails for unattended and autonomous operation}
\label{sec:guardrails}

When an experienced researcher operates the microscope, many mistakes are caught simply because the researcher is watching. In an unattended run, or one in which the next step is chosen by an algorithm or an agent, the onus of protecting the microscope and avoiding recording something that did not happen lies with the automation package. Keeping this in mind, NIMBUS does not rely on a single safety layer for this. It follows a small number of simple rules, each aimed at a particular way that an unattended run can go wrong.

First, some operations are left to researchers and completely off limits to the software package. For example, NIMBUS does not change the high tension, run vacuum cycles, operate the sample autoloader, refill the cold trap, override the beam shutter, insert the EDS detectors or switch the instrument mode in the middle of an acquisition. This list was compiled during live testing on the different microscopes and documented in the repository. 

Second, NIMBUS does not change the state of the column as a hidden side effect of an acquisition. An acquisition command does not lower the viewing screen or switch the microscope into diffraction mode on its own. NIMBUS instead reads the current state and refuses to continue if it is wrong. A selected-area diffraction pattern, for example, is refused when the column is not in a diffraction mode, because a pattern taken in an imaging mode would be saved as diffraction and nothing in the file could reveal the mistake afterwards. To obtain a diffraction pattern, the operator or the experiment script must switch the column into a diffraction mode as an explicit step. An experiment that needs a capability the instrument does not have is stopped before it starts, as described in the previous subsection. A setting the microscope cannot apply, such as the spot/probe size, is refused rather than silently ignored, since a spectrum acquired with the wrong setting looks just like a correct one. Values outside the range an instrument supports are refused in the same way. 

Third, only one client controls the microscope at a time. Without this, accidental commands from two users connected to the same server would be interleaved, and a colleague's stage movement could land in the middle of someone else's tilt series. The server therefore hands control to one client and refuses commands from others with a message that the instrument is busy. Control is renewed at the start and end of every command, including acquisitions that last an hour or more (such as tilt-series).

Fourth, NIMBUS assumes that runs will sometimes fail and makes sure they fail cleanly. Every experiment runs inside a context that carries out the cleanup steps registered during the run, such as stopping a live scan or closing the data store, whether the run finishes, stops on an error or is interrupted by the user. Each cleanup step is attempted on its own, so one failing step does not prevent the others, and the original error is reported rather than hidden by an error raised during cleanup. Failures reported by the vendor interfaces are also checked explicitly, because they do not always appear as errors. Since standard microscope APIs do not set a time limit on their connections, NIMBUS adds one, so that an unresponsive microscope produces an error instead of a run that waits indefinitely.

Rules that depend on the instrument are kept with the code for that instrument rather than in the shared experiments, and each kind is added in a simple, fixed way. A setting the instrument cannot apply is refused by a short check in its client before the request is sent. A parameter limit is stored as a single value in the server for that instrument. Hardware that an acquisition requires is named in the acquisition itself and compared with what the instrument reports when the client connects. Adding a rule therefore means editing only the module for that instrument, and the experiments themselves remain unchanged. Several of the rules above were added in this way after a non-critical problem was first observed during testing on the microscope. We expect further rules to be added as NIMBUS is tested at more sites, together with the user community, making it more resilient over time.

\section{Automated acquisitions on multiple instruments}

To demonstrate cross-platform compatibility using the NIMBUS API, we ran the same high-level code, with instrument specific settings, on a JEOL JEM-F200 at the Max Planck Institute for Solid State Research (MPI-FKF) and a TFS ThemIS at the Molecular Foundry, Lawrence Berkeley National Laboratory. We chose two workflows that differ in complexity and length of the experiment. The first acquires correlated imaging, spectroscopy, and diffraction from a single region of a specimen. The second is an automated electron tomography tilt series on the TFS ThemIS. 

\subsection{Automated electron diffraction and energy-dispersive point spectrum acquisition}

\begin{figure*}[h]
\centering
  \includegraphics[width=\textwidth]{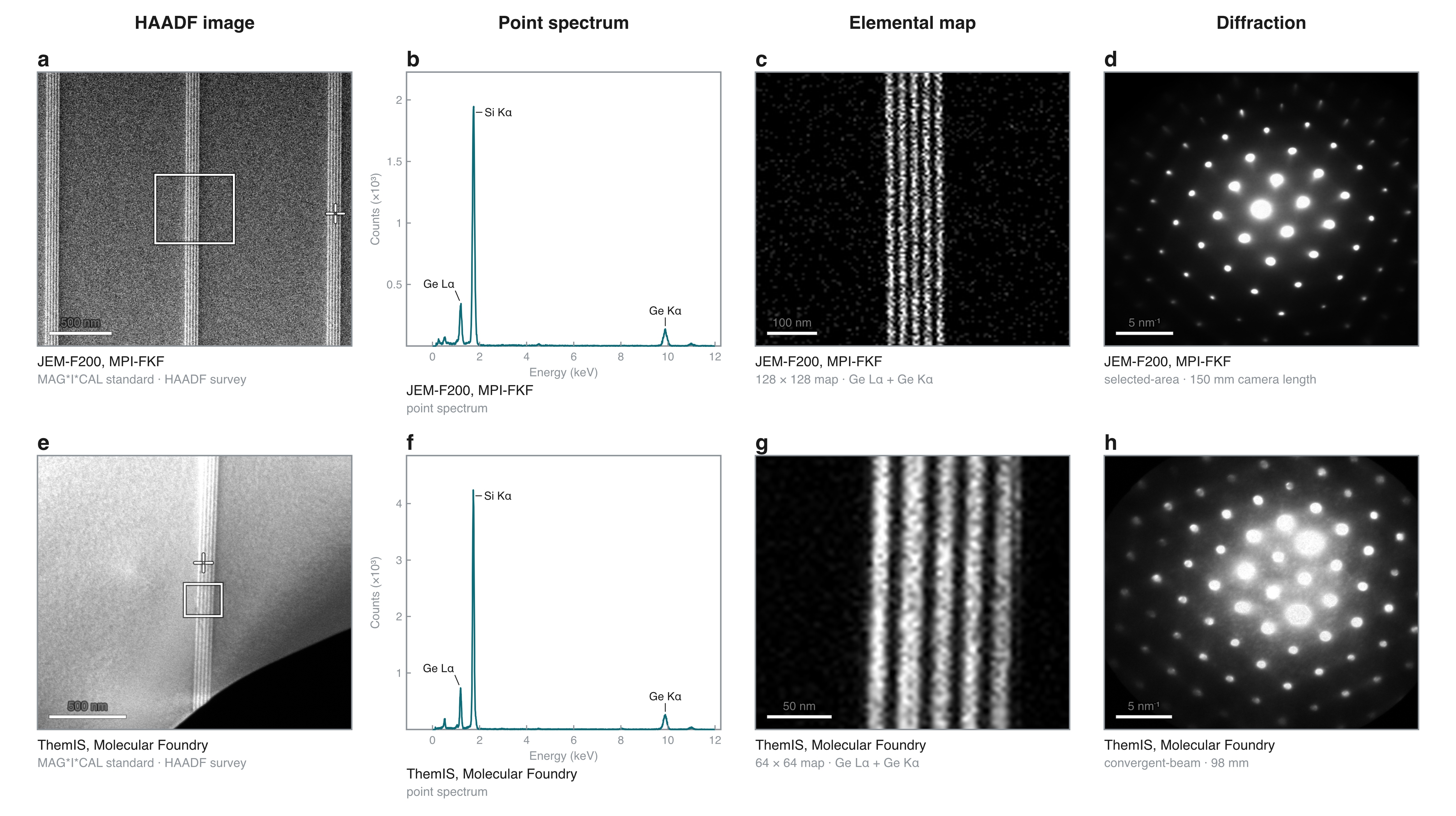}
  \caption{Acquisitions on two instruments of different vendors through the same NIMBUS workflow, arranged by modality. (a--d) JEOL JEM-F200 at MPI-FKF; (e--h) Thermo Fisher ThemIS at the Molecular Foundry, LBNL; both on a MAG*I*CAL Si/Ge calibration standard. (a,e) HAADF survey; the box marks the mapped region and the crosshair the point-spectrum position, both placed from coordinates recorded in the corresponding archives. (b,f) Point X-EDS spectra in raw counts per channel (10~eV and 5~eV channels respectively; peak heights are not directly comparable), showing Si K$\alpha$ with Ge L$\alpha$ and Ge K$\alpha$. (c,g) Elemental maps ($128 \times 128$ and $64 \times 64$) integrated over the Ge L$\alpha$ and K$\alpha$ windows, resolving the individual Ge-bearing layers. (d) Selected-area diffraction at a nominal 150~mm camera length; (h) convergent-beam diffraction at the spectrum position.}
  \label{fgr:demonstration}
\end{figure*}

Characterizing a sample often requires data from several modalities: for example, a HAADF-STEM image for real-space structure, X-EDS for composition, and diffraction for crystallographic information. Acquiring all three signals from the same region manually is cumbersome and requires the operator to click through the vendor software, control the beam by hand, and insert and retract detectors. 

Here we demonstrate a correlative experiment built by chaining NIMBUS API commands, run on instruments from two vendors. The workflow first acquires a HAADF survey of the field of view, on which the user selects regions of interest. It then acquires a point spectrum at each position, an elemental map of the region, and diffraction patterns at the spectrum positions without further input.

On the JEM-F200 at MPI-FKF, we performed this workflow on a MAG*I*CAL standard, producing four NXem archives (survey, point spectrum, elemental map and diffraction) over 59 minutes (Fig.~\ref{fgr:demonstration}a--d), comprising:

\begin{itemize}
    \item Point spectra at a Ge-bearing layer.
    \item A $128 \times 128$ pixel elemental map over a 625~nm field.
    \item A selected-area diffraction pattern at 150~mm nominal camera length.
\end{itemize}

\noindent A similar workflow was then run on the ThemIS on a MAG*I*CAL standard (Fig.~\ref{fgr:demonstration}e--h), comprising:

\begin{itemize}
    \item A point spectrum at a single position.
    \item A $64 \times 64$ pixel elemental map over a 242~nm field.
    \item A convergent-beam electron diffraction pattern at the spectrum position.
\end{itemize}

The same workflow was run on both instruments with instrument-specific settings. All vendor-specific code, PyJEM and FEMTUS on one instrument and TEM Scripting and TIA on the other, is handled by the NIMBUS server modules, so the experiment itself never calls a vendor interface. The JEM-F200 archives were written during acquisition. On the ThemIS, the data were acquired through the NIMBUS API from a jupyter-notebook, and the NXem archives were assembled after the session using the NIMBUS record format. High tension, camera length, convergence angle, probe current, imaging mode, scan rotation and defocus were supplied by the operator afterwards, and each archive records this provenance, including these operator-supplied fields, explicitly. All archives underlying Fig.~\ref{fgr:demonstration} are openly available (see Data availability). This experiment demonstrates NIMBUS as a platform for sharing workflows across different microscopes.

\subsection{Automated electron tomography acquisition}
 \label{sec:tomography}
\begin{figure*}[h]
\centering
  \includegraphics[width=\textwidth]{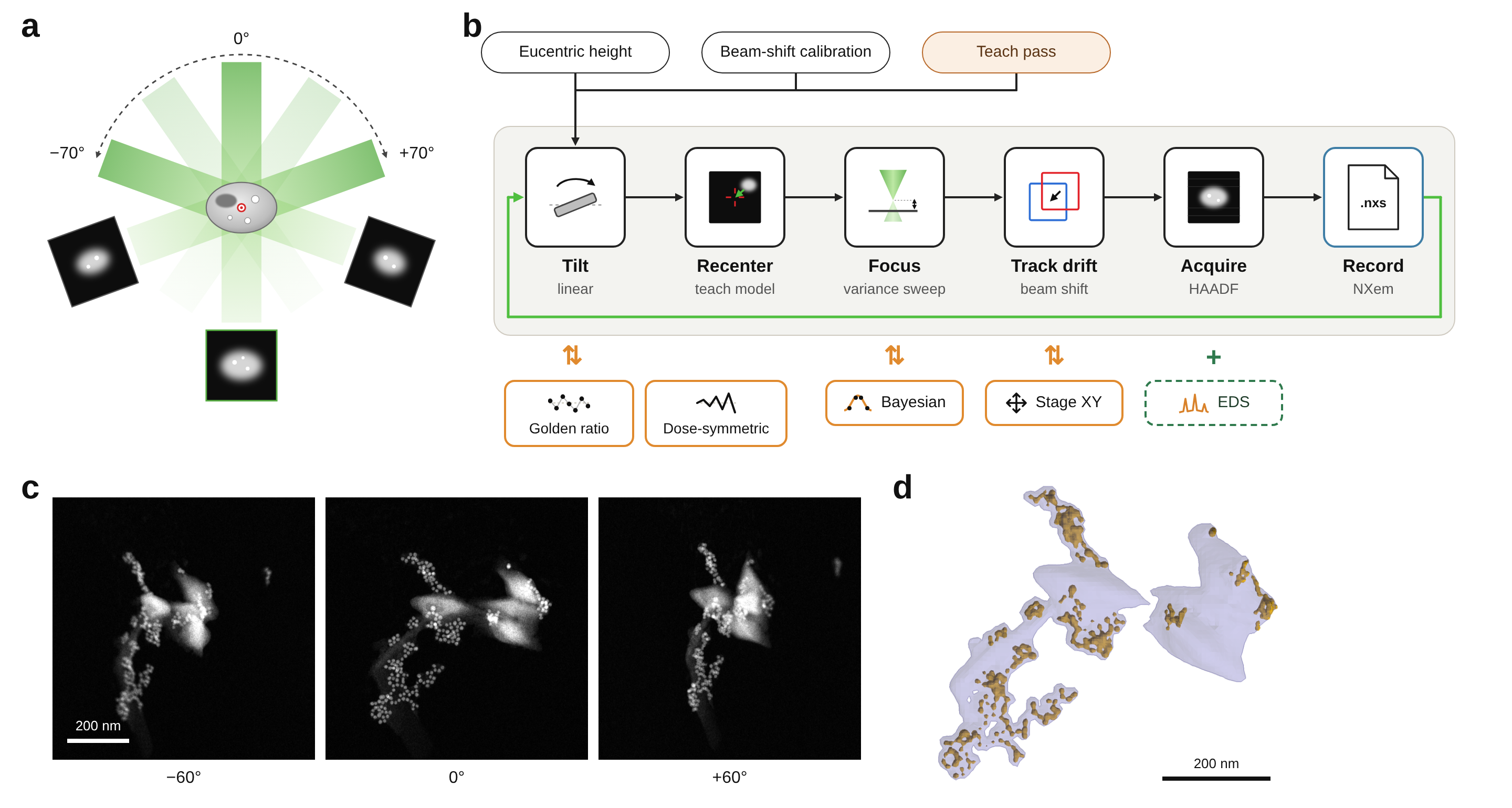}
  \caption{Automated HAADF electron tomography built from reusable NIMBUS steps. (a) Tilt-series geometry from $-70^\circ$ to $+70^\circ$ about the stage $\alpha$ axis. (b) Acquisition loop. Eucentric-height alignment and beam-shift calibration run once before acquisition, with an optional operator coarse tilt-series teach pass. For each projection in the workflow runs tilt, recenter (cross-correlation and beam shift), focus (variance sweep), and finally the final high-quality HAADF acquisition with a NXem recording. Orange boxes are interchangeable implementations of a step; the dashed box is an optional added step. (c) Projections at $-60^\circ$, $0^\circ$ and $+60^\circ$ from the 29-frame series of Te nano-torpedoes with Au fiducials acquired on the Thermo Fisher ThemIS at the Molecular Foundry, LBNL. (d) Segmented reconstruction using an automatic differentiation reconstruction method; Te in purple and Au particles in gold.}
  \label{fgr:tomography}
\end{figure*}

Electron tomography reveals the 3D structure of nanomaterials by acquiring a series of projections as the sample is tilted over a wide angular range, typically $\pm60^\circ$, and computationally reconstructing a volume through an algorithm (Fig.~\ref{fgr:tomography}a). The reconstruction assumes that every projection shows the same region of the specimen with the same focus, viewed only from a different angle. Upon acquiring this data set manually, the operator must ensure that this assumption is enforced in all images. Because microscope stages are mechanically limited in their ability at holding perfect eucentricity and because of stage hysteresis, the region of interest shifts after each tilt and must be recentered. Its height relative to the focal plane also changes with tilt, requiring the operator to refocus. Repeating these corrections for a standard dataset, 50 or more projections depending on the tilt step, takes hours, makes the quality of the series depend on the operator's ability to be concentrated for long lengths of time, and is dose inefficient. Existing automated tomography packages~\cite{mastronarde2005automated} and vendor-supplied routines automate this workflow, but their steps are implemented within each package and cannot easily be modified or reused in other experiments.

In NIMBUS, we developed a general electron tomography acquisition method that is composed of sub-workflows that are available in the NIMBUS API (Fig.~\ref{fgr:tomography}b). We have also developed pre-acquisition workflows such as auto-eucentric height optimization, beam-shift calibration for recentering the sample, and an optional teach-pass to build a model of stage position versus tilt. The acquisition loop then repeats the following six steps at each tilt: tilt the stage, coarse stage recenter, focus with a two-stage variance sweep, perform a cross-correlation computation then use beam-shift to align, acquire the HAADF projection, and record the NXem data file.

As shown in Fig.~\ref{fgr:tomography}, each component of the workflow can be easily modified and replaced. We have also implemented a Bayesian optimization routine~\cite{pattison2025beacon} for dose-efficient auto-focusing. Different tilt sampling schemes, such as dose-symmetric or golden-ratio ordering, can also be used in place of a linear tilt series. Steps can also be added to the acquisition loop, for example, X-EDS spectra can be acquired at specific tilt angles for multimodal electron tomography.

We use this routine to collect an electron tomography dataset of Te particles shaped like torpedoes with Au fiducial nanoparticles acquired using the Thermo Fisher ThemIS at the Molecular Foundry, LBNL, shown in Fig.~\ref{fgr:tomography}c, d. The details of this sample preparation and properties has been published previously by Ben-Moshe \textit{et al.}~\cite{ben2021chain}. The series was acquired with NIMBUS v1~\cite{mishra2026nimbusv1}, which already ran this loop automatically but did not yet write NeXus. After performing calibration and modeling steps, the 29 projections from $-70^\circ$ to $+70^\circ$ in $5^\circ$ steps were acquired without operator intervention, and the series was converted afterwards into an NXem archive that records the commanded tilt of each frame and lists the stage position and the column settings that NIMBUS v1 was able to read and archive. However, the $-70^\circ$ and $+70^\circ$ projections were unusable due to shadowing, and the effective tilt range used for the reconstruction is $-65^\circ$ to $65^\circ$ with 27 projections. The volume in (Fig.~\ref{fgr:tomography}d) is shown and reconstructed using an automatic-differentiation reconstruction algorithm to compensate for the missing wedge artifact~\cite{lim2025missing}.

\section{Data federation and management}

In an automated workflow the record left behind is extremely critical. In an automated workflow without an operator in front of the microscope, the exact instrument state, such as which detectors were inserted or why a measurement value is missing, cannot be recalled later. Therefore, it is imperative for the record to carry this information during the acquisition. NIMBUS aims to record this instrument state during the acquisition and keeps the same form at every deployment site. It is important to note that federation of electron microscopes through NIMBUS means that the measurement records produced at different sites share the same schema, an identifier format and a set of binding concepts, and therefore can be read, compared, combined, and deposited without coordination between the different deployment sites. However, NIMBUS itself does not query or link data across the different institutions automatically. 

\subsection{Multiple site deployment}
\label{sec:sites}

NIMBUS is currently deployed on four microscopes at three institutions: a JEOL JEM-F200 at MPI-FKF, Stuttgart, controlled through PyJEM and FEMTUS; a Thermo Fisher ThemIS at the Molecular Foundry, Lawrence Berkeley National Laboratory, controlled through TEM Scripting and TIA; and a Thermo Fisher Talos 120C and Talos F200i at Stanford University, controlled through AutoScript.  

Running one codebase at several sites requires that nothing site-specific is stored in the code. Every site-dependent value, including the institute code, instrument location, data directory, microscope address, operator identity and material vocabulary, is supplied through the environment or through files maintained by the laboratory. Keeping this in mind, values that necessarily differ between installations have no defaults, since a default could be correct at only one site. Supporting a new instrument requires adding a client and a server module and leaves the site configuration unchanged.

Furthermore in a typical electron microscopy experiment, specimens also travel between different sites with different microscope capabilities or availability. Therefore, it is imperative for the record to distinguish where a sample was prepared from where it was measured. Sample identifiers take the form \texttt{<institute>-<material>-<yyyymmdd>-<nn>}, where \texttt{<institute>} carries the codename of the institute that prepared the sample. Therefore, the measuring institute and the instrument location are recorded separately in every archive. A specimen prepared at MPI-FKF and measured at Berkeley therefore keeps its original identifier while its archive names Berkeley as the place of measurement. This enables future scale-up when multiple different characterization measurements are performed on the same sample at different locations. 

The versatility of the data infrastructure of NIMBUS can also be shown from the difference in the data management platforms at different deployment sites. MPI-FKF deposits the data and the acquisition records into the Kadi4Mat electronic lab notebook (ELN), whereas the Berkeley and Stanford installations operate without one. The archives are nevertheless identical in structure at all three sites, because they are complete before any deposition and their identifiers derive from the sample and run rather than from a catalog. The details of the data deposition to ELNs are described in section~\ref{sec:eln}.

\subsection{Recording instrument state at acquisition and marking missing measurements}
\label{sec:record_instrument}

A conventional electron microscopy workflow records data in a vendor format and rarely converts the data to a community schema afterwards (Fig.~\ref{fgr:workflow}). When a \textit{post hoc} conversion of the data is carried out to a community schema such a conversion can rename fields and reorganize them into a standard hierarchy, but it can only transform what the vendor file already contains. The information most often missing is the state of the instrument at the moment a signal was acquired, such as the beam configuration at a particular measurement position, whether a detector was inserted, or which scan controller generated the raster. These values are absent not because the target schema lacks a place for them, but because the instrument was not queried while they were valid. Once the instrument state changes during the automated acquisition, they cannot be recovered.

\begin{figure*}[!tp]
\centering
  \includegraphics[width=0.95\textwidth]{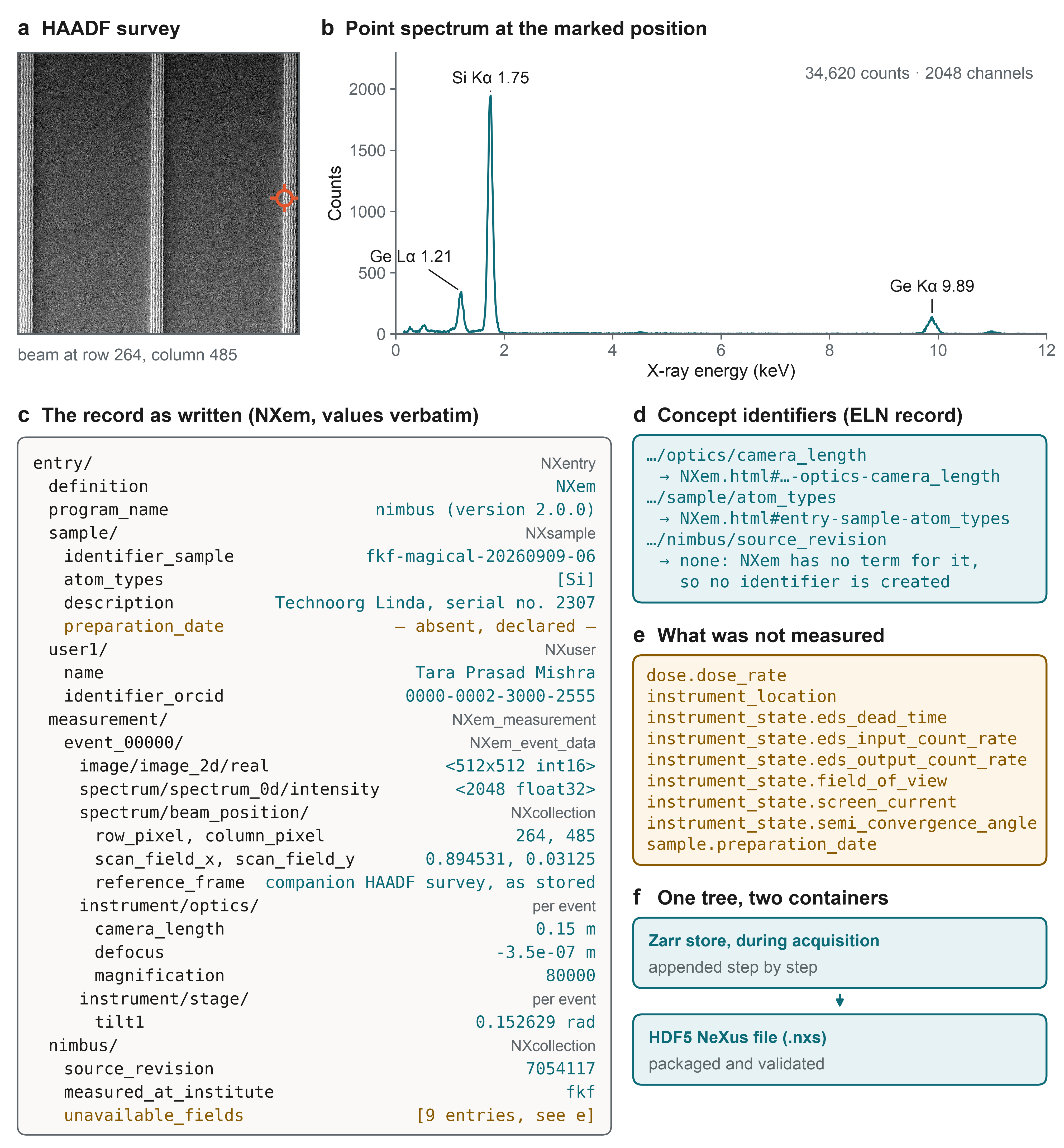}
  \caption{NXem record written by NIMBUS for a point X-EDS acquisition on the JEM-F200. (a) HAADF survey with the beam position (red marker) at which the spectrum was acquired. (b) X-EDS spectrum at that position; silicon K$\alpha$ together with germanium L$\alpha$ and K$\alpha$ from the germanium-bearing layers. (c) The corresponding NeXus/NXem tree, with paths and values as stored. Optics and stage settings are recorded with each acquisition event rather than once per session. (d) In the record deposited in the electronic lab notebook, each field carries the identifier of the NXem concept it represents. In fields where the standard has no term, as for the NIMBUS source revision, no identifier is created. (e) Quantities the instrument could not report are listed explicitly in the record, so a reader can distinguish what was not measured from what was measured as zero. (f) The same tree is held as a Zarr store during acquisition and packaged into a single HDF5 NeXus file afterwards. The sample is recorded as silicon, as declared before the measurement. The element germanium appears only in the measured spectrum, so declared composition and measured evidence are kept apart.}
  \label{fgr:record}
\end{figure*}

As shown in Figure~\ref{fgr:record}, NIMBUS therefore reads the instrument state inside the acquisition loop. Before each signal is acquired (Fig.~\ref{fgr:record}a,b), a single query returns the current state, including the high tension, magnification or camera length, defocus, stage position and tilts, probe configuration and detector state, and this snapshot is written alongside the data it describes under one timestamp. Each event in a run therefore carries its own instrument state (Fig.~\ref{fgr:record}c). Where an acquisition parameter is quantized or adjusted by the instrument, such as the X-EDS energy dispersion or camera exposure times, the value recorded is the one read back from the instrument after setting it, not the one requested. This is primarily due to the fact that sometimes some of the requested values are beyond the range of values that can be set at the microscope or the values are quantized by the microscope itself while setting it. Furthermore, each value is written at its NXem path, so its meaning is fixed by the ratified application definition rather than by a local naming convention. When metadata are extracted from the archive, for example for deposition into a catalog that is not itself schema-aware, every field carries the identifier of the NXem concept it instantiates, so its meaning is universally understandable (Fig.~\ref{fgr:record}d). Where the standard defines no term, NIMBUS does not create one. Such values are placed in a clearly marked local collection and left unbound. One such example is the imaging mode, NXem has no field for it, yet without it a recorded camera length cannot be interpreted. Minting an identifier for a concept that does not exist would prevent a reader from distinguishing local values from standard ones, which is the purpose of binding them.

Although recording the state during acquisition seems straightforward, it makes a second problem quite unavoidable, i.e., not every quantity can be read out reliably from every instrument. For example, during our work at the JEM-F200, the control interface provides no accessor for the electron dose at the specimen. Two simple responses would make the record misleading. On one hand, omitting the field would leave a reader unable to tell whether the value was unavailable, not applicable, or overlooked and on the other, filling it with zero or a nominal value would produce a plausible number that downstream software cannot distinguish from a measurement. NIMBUS instead records such a quantity as unavailable together with the reason, and each archive contains an explicit list of all unavailable fields (Fig.~\ref{fgr:record}e). Derived quantities also follow the same rule. For example, where the beam current is unavailable, the electron dose is not calculated, and the archive states the reason for it. In the event where an operator has supplied a current from an independent measurement (such as a Faraday cup), the value is recorded together with its source, so that it cannot be mistaken for a concurrent measurement.

These measures matter most when no operator is present and a workflow is running autonomously. During an attended session, a person may notice that a value is missing or that a setting did not change. In an automated or agentic workflow, the same value becomes an input to the next decision. A silently substituted zero or an unchecked failure therefore affects not only the record of the current measurement but also the choice of the next one. A record that distinguishes what was measured, what was measured as nothing and what could not be measured allows a decision layer to avoid reasoning from values that do not exist.

\subsection{Hybrid data storage for acquisition and archiving}
\label{sec:storage}
The requirements on a data container during acquisition differ significantly from those on an archive. While a run is in progress, the data must be appendable one event at a time, readable before the run has finished, and robust to interruption. Once the run has ended, the data must form a single, self-contained file that standard tools and the reference validator can read. Due to this constraint, the HDF5 format alone is not suitable as a working container. Although the single-writer/multiple-reader mode of the HDF5 data format allows data to be read while writing, it doesn't allow new groups to be created once writing has started. In the NXem schema, each acquisition event has its own group. The shape of the experiment workflow definition in the automated electron microscopy therefore rules HDF5 out as a live container.

NIMBUS therefore uses two containers for the same record. During the acquisition each run is written to a chunked Zarr working store and then converted into a single NeXus/HDF5 (\texttt{.nxs}) file when it ends, a step we refer to as \textit{packaging} from here on (Fig.~\ref{fgr:record}f). Both are written from one intermediate representation of the NXem tree, so the two containers cannot describe the instrument differently. Furthermore, the Zarr container also fails more gracefully in the event of a crash during the acquisition. The Zarr chunks are independent objects, so an interrupted write affects at most the chunk being written, whereas a partially written HDF5 file can become unreadable in its entirety. NIMBUS writes each acquisition step as a separate event group, following the NXem model, and consolidates the store's metadata only when the run closes. The working store uses the NXem layout but is not itself a NeXus file, because no NeXus specification for Zarr has been ratified yet by the community.

After completing the workflow, packaging copies the arrays into the final file rather than referencing the working store, so the archive remains complete after the store is deleted. Packaging is also repeatable and produces the same file each time. Therefore, a correction to the record, such as a schema fix, can be applied by repackaging the existing store rather than by repeating the acquisition, which matters because instrument time and specimen dose are the costly inputs. Each archive is checked with the reference NeXus validator, and the test suite validates generated archives rather than only testing the code that writes them. 

During the acquisition, the working store also determines what happens when a run is interrupted. The final status of the workflow (completed or incomplete) is recorded explicitly in the archive rather than inferred from the presence of a file or the number of stored frames. An interrupted run is still packaged, but marked as incomplete, so that its completed steps are retained without being mistaken for a finished measurement. As a result, a run interrupted at the 28th of 30 particles still leaves behind a usable archive. An interrupted run can also be resumed from its own store, continuing in the same run and under the same identifier rather than starting a second, partial record. The acquisition recovery process reads the state from the stored data rather than from a separate checkpoint file, so the record of what was completed cannot disagree with the data themselves. Since the working store can be read during acquisition, it can also serve as the shared source of data for processes that monitor a run or select the next measurement, which we return to in Section~\ref{sec:disc_outlook}.

\subsection{Deposition in electronic lab notebooks}
\label{sec:eln}

An electronic laboratory notebook (ELN) is a powerful tool that makes a measurement findable. It enables data to be searched, linked to a sample and a project, and shared with colleagues. However, it does not make the measurement interpretable out of the box. The information needed for interpretation must be captured during acquisition, and information that was not recorded then does not exist to be deposited later. In NIMBUS, deposition is therefore the last step and never a prerequisite. The archive is written before any catalog record is created, and no part of the acquisition, recording or validation code imports a catalog client. This separation is enforced by the test suite. Deposition is disabled unless a catalog host is configured. A laboratory without an ELN, such as the Berkeley and Stanford installations in this work (described in Section~\ref{sec:sites}), obtains archives in the same form as one with an ELN. Where a catalog is configured but unreachable, for example during maintenance or migration, deposition fails with a warning for a run that has already succeeded, and the deposition could be repeated later.

Each deposit is a record, not only an uploaded file. It is built by walking the NXem tree written for the run, rather than from a separately maintained list of fields. This ensures that a field that is no longer written cannot remain silently listed in the record. Every scalar recorded during acquisition is transferred together with its unit and the identifier of the NXem concept it represents. These bindings make it possible to use ELNs that are not schema-aware. Such ELNs accept whatever metadata structure they receive, so the meaning of the deposited record must come from the acquisition format rather than from the ELN. The NeXus file is attached to the record, or referenced by location, size and checksum when it is too large to store directly, and a preview image is attached so that a record can be identified without downloading the archive. Because the record is rebuilt from the data stored on disk rather than from anything held in memory, the same procedure serves both deposition at the end of a run and a repeated deposition weeks later. NIMBUS is therefore agnostic to the choice of ELN. The archive itself can be stored in any repository and remains interpretable without NIMBUS, and any ELN that accepts client-chosen identifiers and nested metadata with units and concept identifiers can be connected through a new backend, whether or not it is itself aware of the NXem schema. Schema-aware systems such as NOMAD~\cite{scheidgen2023nomad}, which read NeXus files directly, could additionally validate and index the deposited records.

At MPI-FKF, NIMBUS deposits into Kadi4Mat~\cite{brandt2021kadi4mat}. Kadi4Mat is one implementation of a catalog interface with five operations: create or update a record, retrieve a record, link two records, list the links of a record, and attach a file. Record creation is idempotent, meaning that repeating it has the same effect as doing it once. Depositing the same run twice, or depositing it again after the archive has been repackaged, therefore updates the existing record rather than creating a second record that also claims to describe the acquisition. A list of operations alone does not guarantee correct behavior. A backend could, for example, create a new record on every deposit or accept links without storing them, and still provide all five operations. NIMBUS therefore defines a common set of tests that every backend must pass. These tests check that depositing the same run twice updates a single record, that records are found by identifiers derived from the sample and run rather than by numbers assigned by the notebook, and that links between records can be read back after they are created. A simple backend that holds records in memory passes the same tests, so the deposition logic can be tested without access to a notebook or a network. Supporting another ELN therefore requires only implementing these five operations and passing these tests.

\section{Discussion}
\label{sec:disc_outlook}

In this work we developed NIMBUS, which is designed as a common foundation for automated electron microscopy, on which both conventional algorithms and agentic workflows can be built. The same workflow runs on JEOL and TFS microscopes, every acquisition leaves a standard record that states what the instrument reported and what it could not, and the guardrails described in Section~\ref{sec:guardrails} apply to every request, whether the next step is chosen by a person, a script or an agent. None of these properties depends on the method used to decide what to measure next, and this separation is what allows the decision-making part of an automated workflow to be changed without changing how the instrument is controlled or how the data are recorded.

Running NIMBUS on real instruments also changed how we approach instrument protection in automated electron microscopy. The usual first instinct is to forbid any operation that carries a risk, but our testing showed that this can create the failures it is meant to prevent. Inserting and retracting the HAADF detector, for example, was initially excluded. However, diffraction on the bottom-mounted camera requires the detector to be retracted, so excluding the motion did not stop it from happening. This artificial safety constraint only meant that this step could not be automated. Beam blanking was also excluded at first, and as a result an image acquired through a blanked beam was stored exactly like a real frame, with nothing in the record to show that it contained only noise. Both operations are now available. The detector can be moved as a step in a workflow, and NIMBUS reads the blanker before an acquisition, unblanks only if necessary and restores the original state afterwards. The rule we have adopted is therefore not to avoid every operation that carries a risk, but to avoid any change to the instrument that NIMBUS cannot see and record. We expect this distinction to matter even more as control is handed to agents, which act on whatever the software reports to them.

With agentic workflows becoming increasingly popular for automation, NIMBUS is designed to serve as the layer between an agent and the microscope, although agentic operation was not evaluated in this work. An agent working through NIMBUS does not need to know the scripting interface of each vendor. It sees a single generalized API with a limited set of operations, and every request passes through the same checks as an experiment written by hand. We expect this to keep the instructions given to an agent short, leaving more of the precious context of a large language model for the experiment itself, and eventually to allow smaller, open-source models to run microscopy workflows. NIMBUS is therefore a natural complement to the agentic frameworks recently developed by the community, such as ADAM \cite{spurgeon2026balancing} and TEM Agent~\cite{wall2026tem}, which give language models access to microscope control and focus on how an agent plans and decides, and which could use NIMBUS as a common instrument layer across JEOL and TFS microscopes. For an agentic workflow, the record is equally important. An agent decides what to measure next from the data it has already seen, and if those data are read from the working store described in Section~\ref{sec:storage}, the archive also shows what information was available at each decision. It also remains to be seen how costly such workflows are in practice, in both instrument time and computation, and whether they can be adopted widely by the community.

While agentic control is a promising direction, deterministic conventional acquisition workflows still have considerable room for improvement. Many of the steps in an automated workflow, such as focusing, recentering and choosing the next tilt angle, are well-defined optimization problems for which better methods can be developed and compared, and machine-learning-driven acquisition has already shown how much can be gained in this way~\cite{kalinin2021automated, roccapriore2022physics,pattison2025beacon}. 
The Bayesian optimization routine for autofocusing mentioned in Section~\ref{sec:tomography} is one such method already available as an interchangeable step in NIMBUS. Because workflows can be replaced individually, a new method can be tested within an existing experiment and compared directly with the one it replaces, on the same instrument and with the same record. 

Finally, manufacturers of modern detectors and scan generators increasingly provide programming interfaces of their own, which make it possible to automate more complex experiments such as four-dimensional STEM and electron energy loss spectroscopy~\cite{pratiush2025hae,pratiush2025orchestrator}. We plan to support these detectors in NIMBUS and to build automated workflows for them, following the same pattern of capabilities, workflows and experiments described here.

\section*{Conclusion}

We have presented NIMBUS, an open platform for automated electron microscopy that controls transmission electron microscopes from multiple vendors (JEOL and TFS) through a single generalized API. A small server on the microscope computer handles the vendor interface, so that a workflow written once on a separate workstation can be run on instruments from different vendors. Instrument-specific guardrails reject unsupported, out-of-range or state-inconsistent requests, complementing rather than replacing vendor interlocks and local safety procedures, and keep an unattended run from recording something that did not happen. Each acquisition is recorded in a standard NXem record that states what the instrument reported and what it could not, and that remains usable after an interrupted run, with deposition into an electronic lab notebook as an optional final step. NIMBUS is currently deployed at three institutions. By separating how the next measurement is chosen from how the instrument is controlled and how the data are recorded, it provides a common foundation for both conventional algorithms and agentic workflows in electron microscopy.

\section*{Author contributions}
T.P.M. conceived the project, designed the architecture and supervised the project. T.P.M. and C.L. implemented the software and wrote the original draft. T.P.M. acquired the JEM-F200 data, prepared the figures and curated and deposited the data. C.L. acquired the ThemIS data, including the tomography series, and reconstructed the tomogram. A.J.P. and P.E. contributed to the initial discussions and the conception of the project for the Thermo Fisher Scientific instruments. K.B. contributed to the deployment and testing on the ThemIS and edited the manuscript. J.T. implemented the Kadi4Mat integration. K.H. assisted with the acquisitions at the JEM-F200. P.A.v.A. provided access to the JEM-F200 facility and contributed to the review and discussion of the manuscript. C.O. edited the manuscript. T.P.M., M.C.S., C.O. and P.A.v.A. acquired funding. All authors reviewed and approved the final manuscript.

\section*{Conflicts of interest}
There are no conflicts to declare.

\section*{Data availability}
The source code of NIMBUS is available at \url{https://github.com/taramatsci/nimbus} under the GNU General Public License v3.0. The version used in this study is v2.0.0, archived on Zenodo at \url{https://doi.org/10.5281/zenodo.24011877}. Installation, deployment and the list of operations NIMBUS does not perform are documented in the repository (\texttt{README.md}, \texttt{llm.md}, \texttt{DEPLOYMENT.md} and \texttt{SAFETY.md}). A simulated microscope included in the package allows the experiments and the recording pipeline to be run without access to an instrument. An earlier release without the vendor-agnostic control language and data layer, v1.0.0, is archived at \url{https://doi.org/10.5281/zenodo.21369051}~\cite{mishra2026nimbusv1}.

Instrument control relies on vendor software that is not distributed with NIMBUS, such as PyJEM and FEMTUS for JEOL microscopes, and TEM Scripting, TIA and AutoScript for Thermo Fisher Scientific microscopes. These are supplied by the respective manufacturers, and AutoScript requires a separate licence.

The NXem archives underlying Figs.~\ref{fgr:demonstration} and \ref{fgr:tomography} are available in NOMAD at \href{https://doi.org/10.17172/nomad.82pd-yt02}{doi:10.17172/nomad.82pd-yt02}~\cite{mishra2026nomad}. The record contains the four JEM-F200 archives written during acquisition, the three ThemIS archives assembled after the session together with the notebook outputs they were built from, and the tilt series acquired with NIMBUS v1 and converted to NXem afterwards, with its conversion script. Each archive records how it was produced and lists the quantities the instrument did not report, and can be validated with the \texttt{nimbus check-archive} command of the released code.

\section*{Acknowledgements}

The authors acknowledge Dr. Assaf Ben-Moshe and Prof. Paul Alivisatos for providing the Te torpedoes shaped with Au fiducial used for the automated electron tomography in section 3.2. CO and CL acknowledge funding support from the Accelerated Materials Design \& Discovery Program in the Toyota Research Institute.

Work at the Molecular Foundry was supported by the Office of Science, Office of Basic Energy Sciences, of the U.S. Department of Energy under Contract No. DE-AC02-05CH11231, under proposal number MFP-10476.

During the preparation of this work the authors used Claude Opus 5.5 (Anthropic) to assist with drafting and editing the manuscript text, and with checking the consistency of statements in the manuscript against the NIMBUS source code and the deposited data. Claude Code (Anthropic) was also used as a coding assistant during the development of the NIMBUS software. All code was reviewed and tested by the authors and validated on the instruments described. No generative AI tool was used to create, alter or analyse the experimental data or images. After using these tools, the authors reviewed and edited all content and take full responsibility for the content of the publication.



\balance


\bibliography{rsc} 
\bibliographystyle{rsc} 
\end{document}